\documentclass[12pt,a4paper]{article}
\usepackage[margin=2cm]{geometry}
\usepackage[margin=5pt,font=small,labelfont=bf]{caption}
\usepackage{graphicx,amsmath,amssymb,natbib,setspace,
  lineno,xspace,color,subcaption,appendix,authblk,
  mathrsfs,array,multirow,cancel}
\definecolor{darkblue}{rgb}{0.0,0.0,0.6}
\usepackage[colorlinks=true,breaklinks=true,linkcolor=darkblue,
            urlcolor=darkblue,anchorcolor=darkblue,citecolor=darkblue]{hyperref} 

\newcommand{\pdiff}[2]{\dfrac{\partial{#1}}{\partial{#2}}}
\newcommand{\Div}{\boldsymbol{\nabla}\cdot}
\newcommand{\Grad}{\boldsymbol{\nabla}}
\newcommand{\sigmabar}{\overline{\boldsymbol{\sigma}}}
\newcommand{\omp}{(1-\phi)}
\newcommand{\permtensor}{\mathbf{K}}
\newcommand{\kperp}{k_\perp}
\newcommand{\kpara}{k_\parallel}
\newcommand{\rxi}{r_\zeta}
\newcommand{\kap}{\boldsymbol{\kappa}}
\newcommand{\cmp}{\mathcal{C}}
\newcommand{\smallpar}{\epsilon}
\newcommand{\xvec}{\boldsymbol{x}}
\newcommand{\yvec}{\boldsymbol{y}}

\newcommand{\vs}{\boldsymbol{v}^S}
\newcommand{\vl}{\boldsymbol{v}^L}
\newcommand{\vvec}{\mathbf{v}}
\newcommand{\sdot}{\dot{s}}
\newcommand{\sdotMax}{\sdot^*}
\newcommand{\thetaMax}{\theta^*}
\newcommand{\dperp}{\delta^*}

\title{Melt-preferred orientation, anisotropic permeability, and
  melt-band formation in a deforming, partially molten aggregate}
\author{Jesse Taylor-West$^1$ and Richard F.~Katz$^2$} \affil{$^1$
  Department of Applied Mathematics and Theoretical Physics,
  University of Cambridge, Cambridge, United Kingdom} \affil{$^2$
  Department of Earth Sciences, University of Oxford, South Parks
  Road, Oxford OX1 3AN, United Kingdom} \affil{email:
  richard.katz@earth.ox.ac.uk}

\begin{document}
\small
\normalsize

\maketitle


\begin{abstract}
  Shear deformation of partially molten rock in laboratory experiments
  causes the emergence of melt-enriched sheets (bands in
  cross-section) that are aligned at about $15$--$20^\circ$ to the
  shear plane.  Deformation and deviatoric stress also cause the
  coherent alignment of pores at the grain scale.  This leads to a
  melt-preferred orientation that may, in turn, give rise to an
  anisotropic permeability. Here we develop a simple, general model of
  anisotropic permeability in partially molten rocks. We use
  linearised analysis and nonlinear numerical solutions to investigate
  its behaviour under simple-shear deformation. In particular, we
  consider implications of the model for the emergence and angle of
  melt-rich bands.  Anisotropic permeability affects the angle of
  bands and, in a certain parameter regime, it can give rise to low
  angles consistent with experiments. However, the conditions required
  for this regime have a narrow range and seem unlikely to be entirely
  met by experiments.  Anisotropic permeability may nonetheless affect
  melt transport and the behaviour of partially molten rocks in
  Earth's mantle.
\end{abstract}

\section{Introduction}
The dynamics of partially molten regions of the mantle are not well
understood; in part, this is due to the inaccessibility of these
regions to direct observation. It is widely agreed, however, that
laboratory experiments performed on synthetic, partially molten
aggregates of olivine and basalt \citep[e.g][]{holtzman03a, king10}
are an effective test of theoretical models of magma dynamics
\citep[e.g.][]{mckenzie84}.  One feature of the experiments that
remains challenging to reproduce in models is the emergence of
high-porosity, melt-rich bands at angles between $15^{\circ}$ and
$20^{\circ}$ to the shear plane \citep{king10}.  Although some
characteristics of the band pattern are sensitive to the applied
strain rate \citep{holtzman07}, the low angles of these bands are
robust to variation in experimental parameters. Here we consider
whether the low angles could arise as a consequence of a permeability
that is directionally dependent (i.e.~anisotropic).

Anisotropic permeability could be a consequence of another empirically
known feature of partially molten rocks subjected to deviatoric
stress: the microstructural alignment of interconnected pockets of
melt between solid grains. This is called melt-preferred orientation
(MPO) and has been observed in many laboratory studies
\citep[e.g.][]{bussod91, daines97, takei10}.  The alignment may be
attributed to the instantaneous state of deviatoric stress
\citep{daines97, takei09a}, or to the combined effects of finite
strain, lattice-preferred orientation, and anisotropic surface energy
of olivine grains \citep{bussod91, daines97, jung98}; it is likely
some combination of the two.  Since the Darcian permeability of
partially molten rocks arises from the shape and interconnectedness of
melt pockets at the grain scale \citep[e.g.][]{bear72, scheidegger74},
it is reasonable to assume that the anisotropic alignment of pores
between grains leads to anisotropy in permeability.  \cite{daines97}
estimated this anisotropy as a function of differential stress and
found that permeability in the direction parallel to the maximum
compressive stress $\sigma_1$ was enhanced by a factor of up to 15
over that parallel to the direction of maximum tensile stress. This is
consistent with a theoretical model for anisotropy of permeability due
to MPO by \cite{hiermajumder11}.

Since both melt-banding at the macroscopic scale and melt-preferred
orientation at the microscopic scale emerge under the same physical
conditions, it is logical to ask whether their dynamics are linked. In
particular, the question we address here is whether the low angle of
high-porosity bands observed in experiments \citep[see schematic
diagram in][fig.~1]{takei13} can be explained through consideration of
anisotropic permeability arising from MPO.  We therefore develop and
analyse two-phase models based on viscous compaction theory
\citep{mckenzie84}. \cite{stevenson89} showed that for a matrix
viscosity that weakens with porosity, unstable growth of porosity
perturbations can lead to localisation of melt. We adopt this rheology
and modify the two-phase theory with assumptions of how deviatoric
stress might modify the permeability of the aggregate.  Calculations
show that anisotropic permeability does indeed exert control over the
predicted angle of melt-rich bands; it can even give rise to the low
angles observed in experiments.  However, the parametric conditions
required to reproduce the angles in experiments are a rather
restrictive set, making it unlikely that this is the true explanation
of the observations.

Other theoretical approaches have been made to explain the emergence
and angle of melt-rich bands. In all cases besides the present one,
authors have introduced more complicated rheological formulations to
obtain low band angles. \cite{katz06} found that low angles are
predicted under non-Newtonian viscosity, but experiments by
\cite{king10} produced low-angle bands and had a measured stress
exponent of $\sim$1.  A more recent approach considers the effect of
deviatoric stress and MPO on diffusion-creep (Newtonian) viscosity
\citep{takei09a, takei09b}. Melt at the grain boundaries is a fast
pathway for diffusion of the component that makes up solid grain;
coherently aligned melt pockets thus give rise to anisotropy of
aggregate viscosity under diffusion creep \citep{takei09a}. The
consequences of microstructural and viscous anisotropy for melt-band
formation were investigated by \cite{takei09c}, \cite{butler12},
\cite{takei13}, \cite{katz13}, \cite{allwright14}, and \cite{takei15}.
All of these studies predict low-angle, high-porosity bands,
consistent with experimental results, but none of them address the
implications of MPO for permeability and melt segregation.

The manuscript is organised as follows. In \S{\ref{sec:governing}} we
present the governing conservation equations, the constitutive laws
and a useful rescaling. \S{\ref{sec:permeabilitytensor}} presents the
assumptions and formulation of the tensorial permeability. In
\S{\ref{sec:LinStab}} we develop the linearised stability
analysis. This section contains sub-sections considering the effect of
the direction of MPO (\S{\ref{sec:angle_effect}}), wavelength of
porosity perturbations (\S{\ref{sec:KappaDep}}), and the anisotropy
conditions that can give rise to low-angle bands
(\S{\ref{sec:low-angles}}). Results from numerical models are
presented in \S{\ref{sec:numerical}} and compared with results from
the stability analysis. In \S{\ref{sec:discussion}} we discuss, summarise,
and conclude the manuscript.

\section{Governing equations}
\label{sec:governing}

\subsection{Conservation statements}
\label{sec:conservation}

Equations describing the dynamics of an aggregate of solid mantle and
liquid magma have been derived by various authors
\citep[e.g.][]{mckenzie84, bercovici01a, rudge11}.  Here we consider a
simplified form of the equations that is relevant to laboratory
experiments.  In this case there is no melting and deformation is
rapid enough that we can neglect the effect of body forces.  The
equations are written in terms of liquid-volume fraction $\phi$,
liquid velocity $\vl$, liquid pressure $p^L$, and matrix velocity $\vs$ as
\begin{subequations}
  \label{eq:AllFour}
  \begin{align}
    0&=\partial_t{\phi}+\Div\left[\phi\vl\right],\\
    0&=\partial_t{(1-\phi)}+\Div\left[\omp\vs\right],\\
    0&=\phi\left(\vl-\vs\right) + 
    \left(\permtensor/\mu\right)\Grad p^L, \label{Darcy}\\
    \label{eq:stokes}
    \boldsymbol{0}&=\Div\sigmabar,
  \end{align}
\end{subequations}
where $\mu$ is the liquid viscosity,
$\sigmabar = \phi\boldsymbol{\sigma}^L + \omp\boldsymbol{\sigma}^S$ is
the stress tensor of the two-phase aggregate, and $\permtensor$ is the
permeability tensor.  These equations represent conservation of mass
for the liquid and solid, and balance of forces for the liquid and the
aggregate, respectively. They incorporate the assumption of constant
and uniform phase densities.

\subsection{Constitutive relations}
\label{sec:constitutive}

To close the system of partial differential
equations~\eqref{eq:AllFour}, we require the specification of
constitutive relations for stress and permeability.

The total stress can be written in terms of the pressure and a
deviatoric stress tensor, $\boldsymbol{\sigma}^L = -p^L\mathbf{I}$ and
$\boldsymbol{\sigma}^S = -p^S\mathbf{I} + \boldsymbol{\tau}^S$, where
we have made the usual assumption that
$\boldsymbol{\tau}^L\approx\boldsymbol{0}$.  Following \cite{rudge11}, we
express the constitutive relationships for compaction and shear as
\begin{subequations}
  \label{eq:constitutive}
  \begin{align}
    p^L-p^S &= \frac{\zeta_\phi}{1-\phi}\Div\vs,\\
    \boldsymbol{\tau}^S &= \frac{\eta_\phi}{1-\phi}\left[\Grad\vs + 
    (\Grad\vs)^T - \tfrac{2}{3}\mathbf{I}\Div\vs\right].
  \end{align}
\end{subequations}
This is consistent with the formulation of \cite{mckenzie84} and gives
\begin{equation}
  \label{eq:bulkstress}
  \sigmabar = -p^L\mathbf{I} + \eta_\phi\left[\Grad\vs + 
    (\Grad\vs)^T - \tfrac{2}{3}\mathbf{I}\Div\vs\right] + 
  \mathbf{I}\zeta_\phi\Div\vs
\end{equation}
as the phase-averaged stress tensor.  $\eta_\phi$ and $\zeta_\phi$ are
the aggregate shear and compaction viscosity.  For present purposes,
it is sufficient to take forms for these that are theoretically
justified \citep{simpson10a, simpson10b} and minimally complicated
\citep{stevenson89}: $\eta_\phi = \eta_0\exp[-\lambda(\phi-\phi_0)]$
and $\zeta_\phi = r_\zeta\eta_\phi$. In the shear viscosity, $\lambda$
controls the reduction in viscosity due to porosity; in the bulk
viscosity, $r_\zeta$ controls the ratio of compaction to shear
viscosity. Parameters $\eta_0$ and $\phi_0$ are constant, reference
values. \cite{stevenson89} showed that the porosity-weakening of
viscosity ($\lambda>0$) gives rise to a melt-banding instability. Here
we take $\lambda=27$, $r_\zeta=5/3$, and $\phi_0=0.05$
\citep[e.g.][respectively]{kelemen97, takei09a, holtzman03a}.

The permeability can be broken down into an isotropic factor,
dependent on the porosity, and an anisotropic factor that we use
to model the effect of melt-preferred orientation.  This gives
\begin{equation} 
  \label{eq:perm}
  \mathbf{K} =  K_0\left(\phi/\phi_0\right)^\ell\mathbf{A},
\end{equation}
where $K_0$ is a reference permeability, $\ell$ is a constant measured
to be between two and three \citep[e.g.][]{vonbargen86, miller14}, and
$\mathbf{A}$ is a non-dimensional, second rank tensor determined by
the form of the permeability anisotropy. Note that $\mathbf{A}$ could
depend on other variables, particularly the deviatoric stress or
finite strain that cause MPO. Here we neglect these dependencies and
take $\ell=3$ for consistency with previous studies.

\subsection{Rescaling}
\label{sec:rescaling}

Equations \eqref{eq:AllFour}, \eqref{eq:bulkstress}, and
$\eqref{eq:perm}$ can be combined to eliminate $\vl$ resulting
in the system
\begin{subequations}
  \label{eq:dimensional_conservation}
  \begin{align}
    \partial_t{\phi} &= \Div \left[(1-\phi) \vvec\right], \label{dimconseq1}\\
    \Div\vvec &= \Div\left[(\mathbf{K}/{\mu})\Grad P\right], \label{dimconseq2}\\
    \Grad{P} &= \Div\left[\eta_\phi(\Grad\vvec+\Grad\vvec^T)\right]+
                 \Grad\left[\left(\zeta_\phi-\tfrac{2}{3}\eta_\phi\right)\Div\vvec\right],
                 \label{dimconseq3}
  \end{align}
\end{subequations}
where we have substituted $P\leftarrow p^L,\,\vvec\leftarrow\vs$ for simplicity
of expression.

We now introduce the following characteristic scales:
$[\boldsymbol{x}]\sim\delta$, $[t]\sim\dot{\gamma}^{-1}$,
$[\eta_\phi]\sim\eta_0$, $[\vvec]\sim\dot{\gamma}\delta$,
$[P]\sim\eta_0\dot{\gamma}$, and $[\permtensor]\sim K_0$. Here we have
used the compaction length $\delta=\sqrt{(r_\zeta+4/3)\eta_0K_0/\mu}$,
a characteristic lengthscale for liquid/solid interaction
\citep{mckenzie84}.  Our focus will be simple-shear flows with
initially uniform strain rate $\dot{\gamma}$, which gives us a
characteristic time-scale for the problem.  Using these scales to
nondimensionalise all symbols we obtain
\begin{subequations}
  \label{eq:nondimensional_conservation}
  \begin{align}
    \partial_t{\phi} &= \omp\cmp - \vvec\cdot\Grad\phi, \label{nondimconseq1}\\
    \cmp &= (r_\zeta+4/3)^{-1}\Div\left[(\phi/\phi_0)^\ell\mathbf{A}\Grad
           P\right], \label{nondimconseq2}\\
    \Grad{P} &= \Div\left[\eta_\phi(\Grad\vvec+\Grad\vvec^T)\right]+
                 \Grad\left[\eta_\phi\left(r_\zeta-2/3\right)\cmp\right],
                 \label{nondimconseq3}
  \end{align}
\end{subequations}
where we have defined the compaction rate as $\cmp\equiv\Div\vvec$.

\section{Formulation of anisotropic permeability}
\label{sec:permeabilitytensor}

\begin{figure}[ht]
  \centering
  \includegraphics[width=16cm]{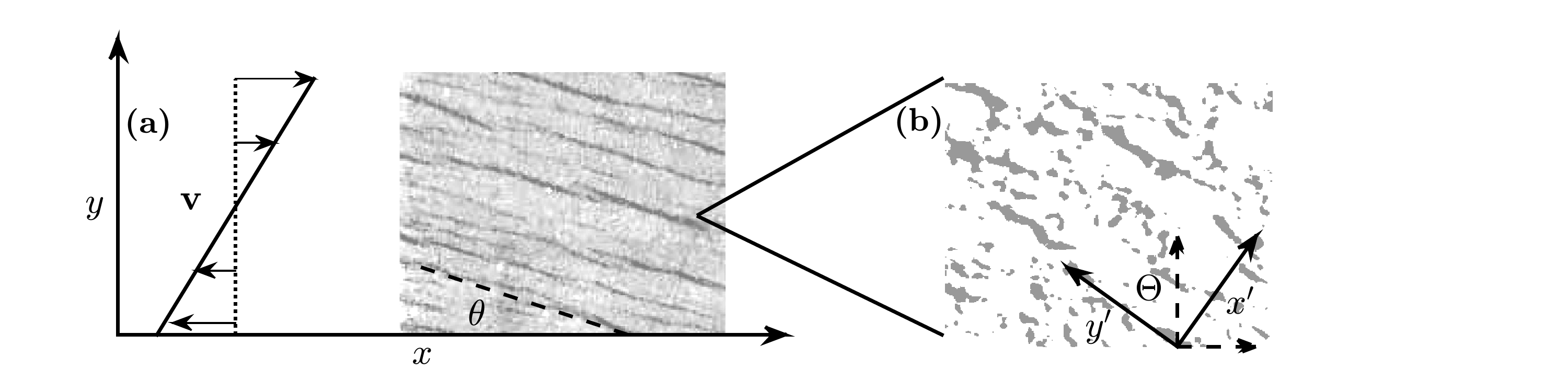}
  \caption{Schematic diagram of melt-rich bands and melt-preferred
    orientation.  \textbf{(a)}~shows a macroscopic image of melt-rich
    bands, aligned at an angle $\theta$ to the shear plane. The image
    is adapted from Figure~4 of \cite{king10}.  A schematic
    illustration of the simple-shear flow field is shown at
    left. \textbf{(b)}~shows a schematic of melt-preferred orientation
    aligned at an angle $\Theta$ to the shear plane. The image is
    adapted from Figure~2 of \cite{holtzman03a} by thresholding the
    greyscale such that melt is grey and solid is white. The
    melt-preferred orientation is aligned with a rotated coordinate
    system $(x',y')$ and has enhanced permeability in the direction of
    the long axes of the melt pockets ($y'$).  The scale of this image
    is microscopic --- of order 10$\times$ the grain size --- so well
    below the continuum scale of high-porosity bands.}
  \label{fig:Shear}
\end{figure}

The dynamic origins of melt-preferred orientation remain poorly
understood.  In laboratory deformation experiments, melt pockets
become elongated such that in cross section, they have a long axis and
a short axis, as shown in Fig.~\ref{fig:Shear}. Coherent alignment
of the long axes of melt pockets results from an unknown combination
of the instantaneous stress \citep{takei09a} and the finite strain
\citep{daines97, jung98}. The latter acts to align olivine grains,
which have anisotropic surface energy and hence anisotropic wetting
affinity.  If the deviatoric stress is the dominant forcing of MPO, we
would expect melt pockets to be in instantaneous alignment, with their
long axes perpendicular to the direction of maximum tension (the
$\sigma_3$-direction).  We can generalise this, however, and
investigate the consequences of MPO alignment in terms of the angle
$\Theta$ between the shear plane and the normal to the long axis of a
melt pocket.

This angle is used to define a rotated coordinate system as in
Fig.~\ref{fig:Shear}. The $x'$-direction is normal to melt pockets and
expected to have a low permeability. The $y'$-direction is parallel to
melt pockets (and in the plane defined by the eigenvectors
corresponding to $\sigma_1$ and $\sigma_3$); this direction is
expected to have a high permeability.

We quantify these expectations with the tensor $\mathbf{A}'$ in the
rotated coordinate system,
\begin{align}
  \mathbf{A'}=\left( 
  \begin{array}{cc} \kperp & 0  \\
    0 & \kpara   \end{array} \right),
\end{align}
where $\kpara\geq 1$ is the permeability enhancement along $y'$ and
$0\leq\kperp\leq 1$ is the reduction along $x'$. A symmetry argument
(Appendix~\ref{sec:diagonal}) shows that $\mathbf{A'}$ must be
diagonal. Then, by rotating from the primed to the unprimed
coordinates, the anisotropy matrix is written as
\begin{align} 
\label{eq:permeability1}
  \mathbf{A}=\mathbf{R}\mathbf{A'}\mathbf{R}^T=\left( 
  \begin{array}{cc}
    \kperp\cos^2\Theta+\kpara\sin^2\Theta 
    & (\kperp-\kpara)\cos\Theta\sin\Theta \\
    (\kperp-\kpara)\cos\Theta\sin\Theta
    & \kperp\sin^2\Theta+\kpara\cos^2\Theta 
  \end{array} \right).
\end{align}
Here we could factor out $\kpara$, for example, and lump it with
$K_0$, reducing by one the number of parameters in the problem.
However, since we have non-dimensionalised lengths with the compaction
length (and hence with $K_0$), this would obscure the role of
permeability enhancement parallel to MPO.

Much of the work below is in understanding the behaviour of solutions
to the system \eqref{eq:nondimensional_conservation} under different
assumptions for the values of $\kperp$, $\kpara$, and $\Theta$ in the
tensor~\eqref{eq:permeability1}.  A list of the key non-dimensional
symbols used in this manuscript is provided in
table~\ref{tab:symbols}. 

\begin{table}[ht]
  \centering
  \begin{tabular}{clll}
    Symbol & meaning & equation, value, or range \\ \hline
    $\xvec$ & position &\\
    $t$ & time & \\
    $P$ & liquid pressure & \\
    $\vvec$ & velocity & \\
    $\phi$ & porosity & &\\
    $\cmp$ & compaction rate & $\Div\vvec$\\
    $\eta_\phi$ & shear viscosity & $\eta_\phi=\exp\left[-\lambda(\phi-\phi_0)\right]$\\
    $\zeta_\phi$ & bulk viscosity & $\zeta_\phi=r_\zeta\eta_\phi$\\
    $\lambda$ & porosity-weakening exponent & 27 \\
    $r_\zeta$ & viscosity ratio & 5/3 \\
    $\phi_0$ & reference porosity & 0.05 \\
    $\ell$ & permeability exponent & 3\\
    $\mathbf{A}$ & permeability anisotropy tensor & eqn.~\eqref{eq:permeability1}\\
    $\kpara$ & MPO-parallel permeability factor & 1--200\\
    $\kperp$ & MPO-perpendicular permeability factor & 0.005--1\\
    $\Theta$ & angle of MPO to shear plane & $0^\circ$--$90^\circ$ \\
    $\theta$ & angle of perturbation wavevector to shear plane 
                     & $0^\circ$--$180^\circ$\\
    $\smallpar$ & very small number & $\ll1$ \\
    $s,\sdot$ & perturbation amplitude, growth rate 
              & eqn.~\eqref{perturbation}, \eqref{eq:growthrate}\\
    $\kap,\kappa$ & perturbation wave-vector, magnitude 
                     & $\kap=\kappa\left(\sin\theta,\cos\theta\right)$\\
    \hline
  \end{tabular}
  \caption{Dimensionless symbols and their meaning.}
  \label{tab:symbols}
\end{table}

\section{Linearised stability analysis}
\label{sec:LinStab}

Equations \eqref{eq:nondimensional_conservation} form a coupled,
non-linear, time-dependent system.  To make analytical progress, we
follow previous workers \citep[e.g][]{spiegelman03b, katz06} and
employ a linearised stability analysis.  We expand the problem
variables in a power-series of $\smallpar\ll 1$ and truncate after
first order,
\begin{equation}
  \label{eq:expansion}
  \left\{\begin{array}{l}
      \phi = \phi_0 + \smallpar\phi_1\left(\xvec,t\right),\\
      P = P_0 + \smallpar P_1\left(\xvec,t\right),\\
      \vvec = \vvec^{\left(0\right)}\left(\xvec\right) + 
      \smallpar\vvec^{\left(1\right)}\left(\xvec,t\right),\\
      \cmp = \cmp_0 + \smallpar\cmp_1\left(\xvec,t\right).
    \end{array} \right.
\end{equation}
Terms of order $\smallpar^0$ are taken to define the
\textit{base-state} about which we perturb.  The expansion of porosity
can be used to express the nondimensional permeability and viscosity
to first order as
\begin{subequations}
  \label{eq:expansion_constit}
  \begin{align}
    \left(\phi/\phi_0\right)^\ell &= 1+\smallpar \ell\phi_1/\phi_0\\
    \textrm{e}^{-\lambda(\phi-\phi_0)} &= 1-\smallpar\lambda\phi_1.
  \end{align}
\end{subequations}
Substituting \eqref{eq:expansion} and \eqref{eq:expansion_constit}
into \eqref{eq:nondimensional_conservation} and requiring terms to
balance at each order allows us to derive two systems of linear
equations at $\mathcal{O}(\smallpar^0)$ and
$\mathcal{O}(\smallpar^1)$. We assume that the leading-order porosity
$\phi_0$ is uniform and that the flow is simple shear with
$\vvec^{(0)} = (\dot{\gamma}y,0)$. The uniform, leading-order
viscosity and permeability therefore result in $\cmp_0=0$ and
$P_0=\textrm{const}$.

The first order balance of equations
\eqref{eq:nondimensional_conservation} becomes
\begin{subequations}
  \label{eq:firstorder}
  \begin{align}
     \partial_t{\phi_1} &= (1-\phi_0)\cmp_1 - \vvec^{(0)}\cdot\Grad\phi_1, \label{timeevoution}\\
    \cmp_1 &= (r_\zeta+4/3)^{-1}\Div\left[\mathbf{A}
             \Grad P_1\right],\label{compaction1}\\
    \Grad P_1 &= \Div\left[\Grad\vvec^{(1)}+(\Grad\vvec^{(1)})^T\right]
                +(\rxi-2/3)\Grad\cmp_1-\lambda\Div\left[\phi_1
                \left(\Grad\vvec^{(0)}+(\Grad\vvec^{(0)})^T\right)\right].
                \label{gradP}
  \end{align}
\end{subequations}
We consider harmonic porosity perturbations that are advected by the
background simple shear 
\begin{equation}
  \label{perturbation}
  \phi_1 = \exp\left[i\kap\cdot\left(\xvec-\int_0^{t}\vvec^{(0)}
      \text{d}t\right)+s(t)\right],
\end{equation}
where $\exp s(t)$ is the time-dependent amplitude and the wave-vector
is given as
$\kap=\left(\kappa_x,\kappa_y\right) =
\kappa\left(\sin\theta,\cos\theta\right)$.
Hence $\theta$ represents the angle between the perturbation wave
vector $\kap$ and the $y$-axis or, equivalently, between the
perturbation wavefronts and the shear plane. 

This linearised system \eqref{eq:firstorder} can be solved, along with
\eqref{perturbation}, to obtain the growth rate (details in
Appendix~\ref{sec:solve})
\begin{equation}
  \label{eq:growthrate}
  \sdot = \lambda\frac{1-\phi_0}{r_\zeta+4/3}
  \sin{2\theta}\frac{\left\langle\kap,
  \mathbf{A}\kap\right\rangle}{1+\left\langle\kap,
  \mathbf{A}\kap\right\rangle},
\end{equation}
where the inner product notation means
$\left\langle\kap, \mathbf{A}\kap\right\rangle =\sum_{i}\limits
\sum_{j}\limits\kappa_i A_{ij}\kappa_j$.

For isotropic permeability $\kperp=\kpara=1$; then from
equation~\eqref{eq:permeability1}, $\mathbf{A}$ reduces to the
identity matrix. In this case, the growth rate is identical to that
calculated by \cite{spiegelman03b} for isotropic permeability (and
viscosity). Further, in the limit of vanishing perturbation wavelength
$\kappa\to\infty$ (where $\kappa=\vert\kap\vert$), the ratio
$\left\langle\kap, \mathbf{A}\kap\right\rangle/(1+\left\langle\kap,
  \mathbf{A}\kap\right\rangle)\to 1$
and so the growth rate again reduces to that of the isotropic case.
In contrast, the limit $\kappa = 0$ has zero growth rate. It can be
shown (Appendix~\ref{sec:solve}) that the greatest sensitivity to the
anisotropy of permeability occurs where $2\pi/\kappa \sim 1$ or, in
dimensional terms, when the wavelength of the instability is
approximately equal to the compaction length.

\begin{figure}[ht]
  \centering
  \includegraphics[width=16cm]{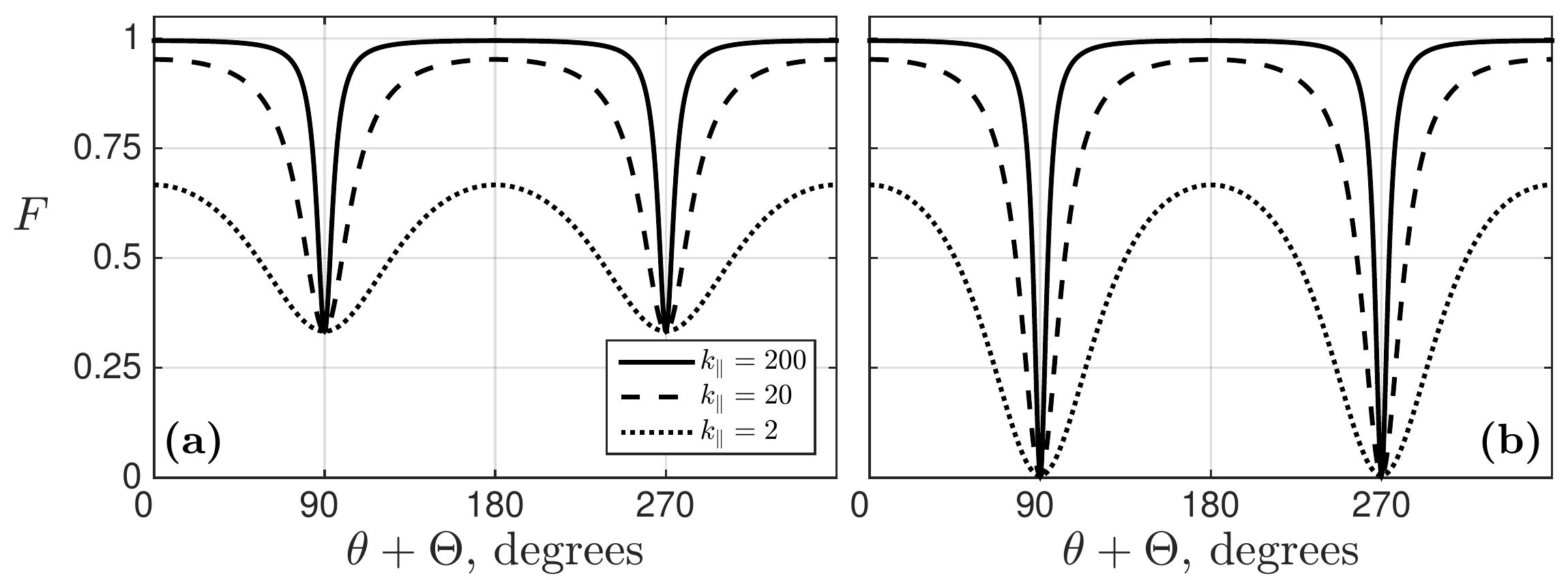}
  \caption{Anisotropy factor $F$ as a function of $\theta + \Theta$
    for $\kappa=1$ and various values of $\kpara$.  
    \textbf{(a)}~$\kperp=0.5$.  \textbf{(b)}~$\kperp= 0.005$. }
     \label{fig:F}
\end{figure}

For a permeability matrix $\mathbf{A}$ from equation
\eqref{eq:permeability1} we can expand and simplify to obtain
$$\left\langle\kap,\mathbf{A}\kap\right\rangle =\kappa^2\left[\kpara -
  (\kpara-\kperp)\sin^2(\theta+\Theta)\right].$$
Hence from equation \eqref{eq:growthrate}, we are interested in the
behaviour of the factor
\begin{equation}
  \label{eq:anisotropyfactor}
  F=\frac{\kappa^2\left[\kpara - (\kpara-\kperp)\sin^2(\theta+\Theta)\right]}
  {1+\kappa^2\left[\kpara - (\kpara-\kperp)\sin^2(\theta+\Theta)\right]}.
\end{equation}
that modifies the isotropic growth rate.  $F$ is plotted in
Fig.~\ref{fig:F} as a function of $\theta+\Theta$ for $\kappa=1$ and
various values of $\kpara$ and $\kperp$. MPO is parallel to porosity
perturbations when $\theta+\Theta = 90^\circ$ or $270^\circ$.  At
these two angles, the reduced permeability $\kperp$ in the
MPO-perpendicular direction controls the flow of melt perpendicular to
bands and causes a minimum in $F$. With $\kperp\to 0$, melt cannot
segregate in the direction perpendicular to MPO, and hence it cannot
feed bands that are oriented parallel to MPO.  When
$\theta+\Theta \neq 90^\circ$ or $270^\circ$, values of $\kpara>1$
facilitate segregation of melt into (or out of) perturbations.

Although permeability can promote or impede melt transport, it does
not drive melt transport.  In this context, it is important to note
that in the linearised analysis, the permeability does not multiply
any zeroth-order fields and hence perturbations to the permeability
play no role in the first-order balance~\eqref{eq:firstorder}. Only
anisotropy of the zeroth-order permeability affects the growth rate of
porosity perturbations. Here we assume that MPO (and hence anisotropic
permeability) emerges on a much shorter time scale than melt
redistribution, so the leading-order MPO is in equilibrium with the
leading-order stress state. This assumption is consistent with
grain-scale models \citep{hiermajumder11} as well as experimental
observations \citep{takei10, zimmerman99}.  Other experiments indicate
that the strength of MPO increases with the amount of time that the
stress is applied \citep{daines97}. Furthermore, the local stress
state is affected by the segregation of melt into bands
\citep[e.g.][]{takei15}, which could modify the MPO and the
permeability.  However, in the context of linearised analysis, this
would not appear at leading order and hence would not modify our
results.

Equation~\eqref{eq:growthrate} shows that the growth-rate of
perturbations depends on the product of $F$ with $\sin2\theta$, where
the latter represents the deviatoric stress (tension positive) that
drives melt segregation.  In the absence of anisotropic permeability,
$\sin2\theta$ is the sole angular dependence of the growth rate and
bands are predicted to grow fastest at $45^\circ$ to the shear plane.
We next explore how and when $F$ modifies this prediction.

\subsection{MPO orientation and growth rate}
\label{sec:angle_effect}

We return to \eqref{eq:growthrate} and consider the role of the
orientation of anisotropy $\Theta$. Within the limitations of
linearised analysis, it is the porosity perturbation with the largest
instantaneous growth rate that is predicted to dominate after finite
time, and hence to be expressed by the physical system.  We are
therefore interested in the angle $\thetaMax$ at which the growth rate
reaches its maximum, $\sdotMax$.  Since the growth rate is most
sensitive to anisotropy at $\kappa\sim1$, we consider only $\kappa=1$
in this section; this constraint is relaxed in the following
sections.

\begin{figure}[h]
  \centering
  \includegraphics[width=.5\linewidth]{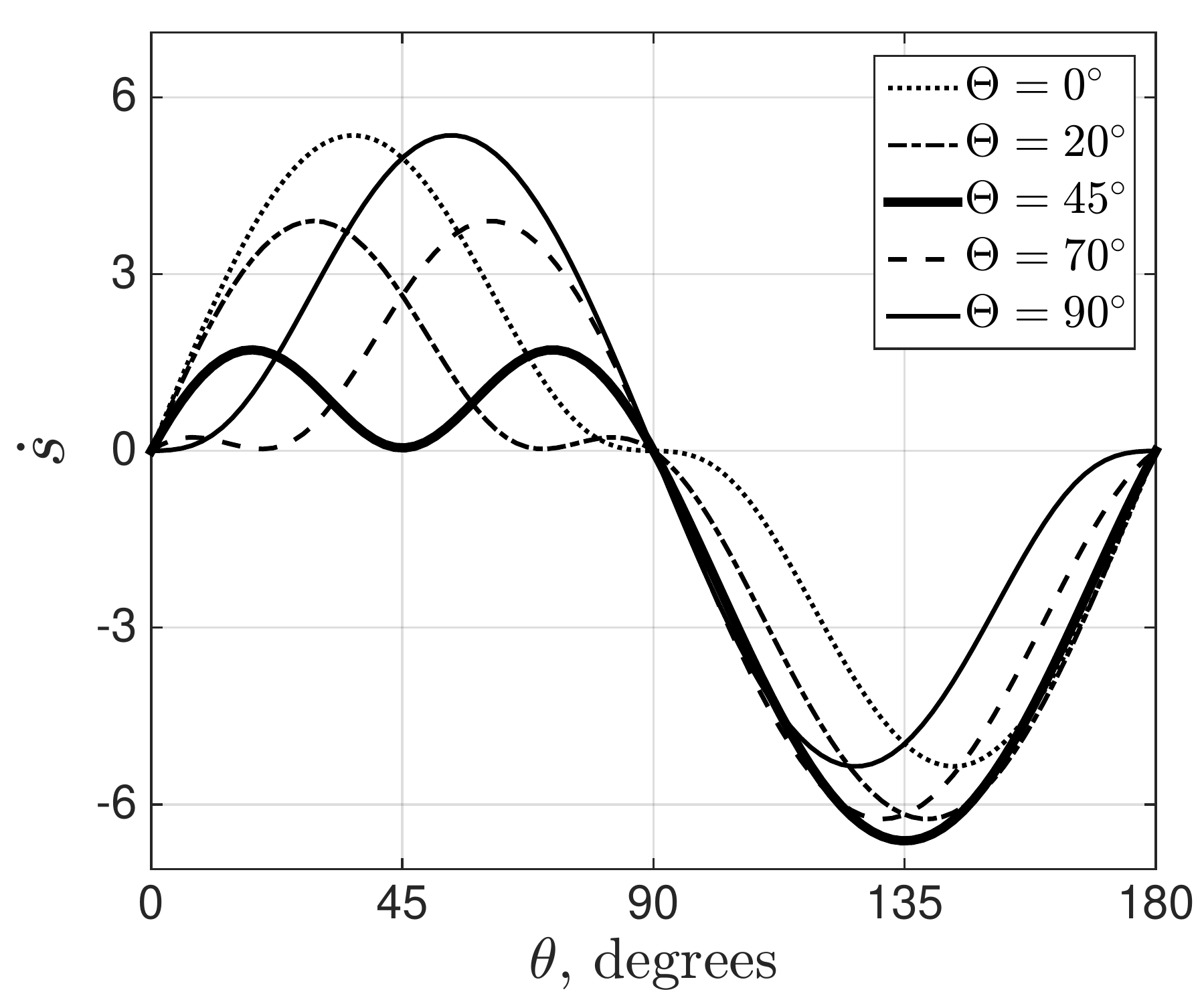}
  \caption{Growth-rate $\sdot$ versus $\theta$ for various values of
    anisotropy angle $\Theta$. Each curve is calculated with
    $\kperp=0.005$, $\kpara=2$.}
  \label{fig:VariousTheta}
\end{figure}

Fig.~\ref{fig:VariousTheta} shows the growth rate versus $\theta$ for
a set of values of anisotropy angle $\Theta$.  Independent of any
imposed anisotropy, perturbation growth is driven by tension across a
high-porosity, viscously weak band \citep{stevenson89}. As band angles
rotate into directions parallel or perpendicular to the shear plane,
this tension vanishes and the growth rate goes to zero
\citep{spiegelman03b}; this is why all curves in
Fig.~\ref{fig:VariousTheta} cross $\sdot=0$ at $0^\circ$ and
$90^\circ$. Anisotropy of permeability modifies the $\sin2\theta$
dependence of growth-rate predicted by \cite{spiegelman03b}. As
$\Theta$ is increased from its smallest value, a local minimum in the
positive growth rates appears in the range
$0^\circ\le\theta\le90^\circ$.  This minimum exists at the angle where
perturbation wavefronts are parallel to MPO,
$\theta = 90^\circ-\Theta$.  For the case of $\Theta=45^{\circ}$, the
minimum occurs at $\theta=45^\circ$ and we find two peaks of equal
growth rate, one at an angle lower than $45^{\circ}$ and one
higher. \cite{katz06} show that although the two perturbation angles
have the same instantaneous growth rate, it is the lower-angle
perturbation that dominates, since the high-angle band is rotated more
rapidly by the background, simple-shear flow out of favourable
orientation for growth (eqn.~\eqref{perturbation}).

\subsection{Perturbation wavelength and growth rate}
\label{sec:KappaDep}

\begin{figure}[h!]
  \centering
  \includegraphics[width=.5\linewidth]{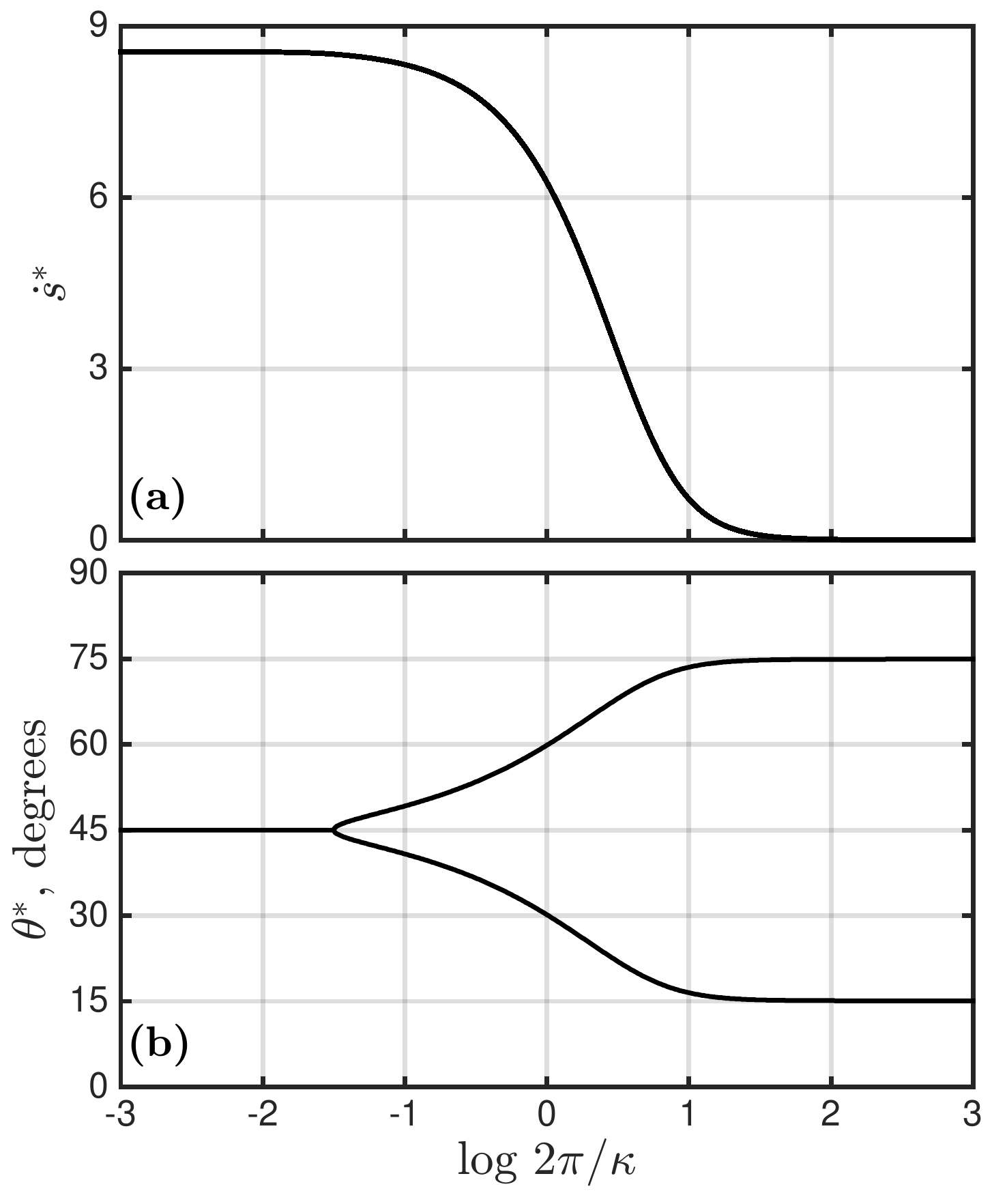}
  \caption{The effect of perturbation wavelength $2\pi/\kappa$ when
    $\Theta=45^{\circ}$, $\kperp=0.005$ and
    $\kpara=2$. \textbf{(a)}~Maximum growth rate, $\sdotMax$, against
    log-wavelength.  \textbf{(b)}~$\thetaMax$ against
    log-wavelength. The angle of both the low- and the high-angle
    growth peaks is plotted.}
  \label{fig:wavelength_effect}
\end{figure}

As in previous work on isotropic permeability
\citep[e.g.][]{stevenson89}, the growth rate in
equation~\eqref{eq:growthrate} depends on the wavelength of
perturbations relative to the compaction
length. Fig.~\ref{fig:wavelength_effect}(a) plots growth rate against
the log of perturbation wavelength for the angle $\thetaMax$ that has
maximum growth rate. Here, we consider only MPO aligned at
$\Theta=45^\circ$. For perturbation wavelengths much larger than the
compaction length, the growth rate of perturbations drops to
zero. Over these distances, pressure gradients associated with
variation in viscosity are small and cannot drive melt segregation and
differential compaction.

Looking instead at wavelengths much smaller than the compaction
length, Fig.~\ref{fig:wavelength_effect}(a) indicates that there is no
wavelength selection by the growth rate.  At vanishingly small
perturbation wavelengths, the resistance to segregation by Darcy drag
is negligible.  This highly local segregation is insensitive to the
permeability; its orientation of maximal growth rate is therefore
determined entirely by the orientation of the stress tensor and hence
by $\sin2\theta$.  Fig.~\ref{fig:wavelength_effect}(b) plots
$\thetaMax$ as a function of perturbation wavelength; it confirms that
at wavelengths smaller than a tenth of the compaction length
$\thetaMax=45^{\circ}$, as in the isotropic case.  For increasing
wavelength, anisotropy of permeability plays an increasing role; the
$45^{\circ}$ growth-rate peaks splits into two peaks of equal height
(as in Fig.~\ref{fig:VariousTheta} for $\Theta=45^\circ$).  However,
this split occurs as the growth rate declines to zero. Therefore, in a
full, nonlinear solution to the governing equations that is
initialised with a broadband spectrum of perturbations, the growth of
shorter-wavelength perturbations would dominate the porosity field.
These would likely appear at $\sim45^\circ$. However \cite{butler10}
showed that the wavelength of high porosity bands in numerical
simulations increases as the system evolves, due to the kinematics of
the background flow. He explained that bands grow, rotate, and widen
until wavelengths are similar to one compaction length.  This is, as
we have shown, the length scale at which anisotropic permeability has
a significant effect on band angles.

It is also important to note, however, that application of the present
model to arbitrarily small wavelength is invalid because the continuum
assumptions hold only at length-scales much larger than the solid
grains.  Indeed if we consider the formation of ``bands'' \textit{at}
the grain scale, the distribution of melt is indistinguishable from
melt-preferred orientation.  So for a variation in porosity to be
considered a melt-rich band, the variation must take place over a
length-scale on which the concept of porosity is well defined; i.e.~a
large multiple of the mean grain size. Such wavelength are, in fact,
observed in experiments \citep{holtzman03a, king10} indicating that
either the initial variations of porosity are concentrated at
intermediate or larger scales, or that the non-linear system is
regularised over finite time by some process that has not been
included in the theory above. Possible examples of the latter are
surface-energy-driven flow \citep{bercovici15} and
dissolution/transport/precipitation due to gradients in chemical
potential \citep{takei09d}.

\subsection{The conditions for low-angle bands}
\label{sec:low-angles}

\begin{figure}[h]
  \centering
  \includegraphics[width=15cm]{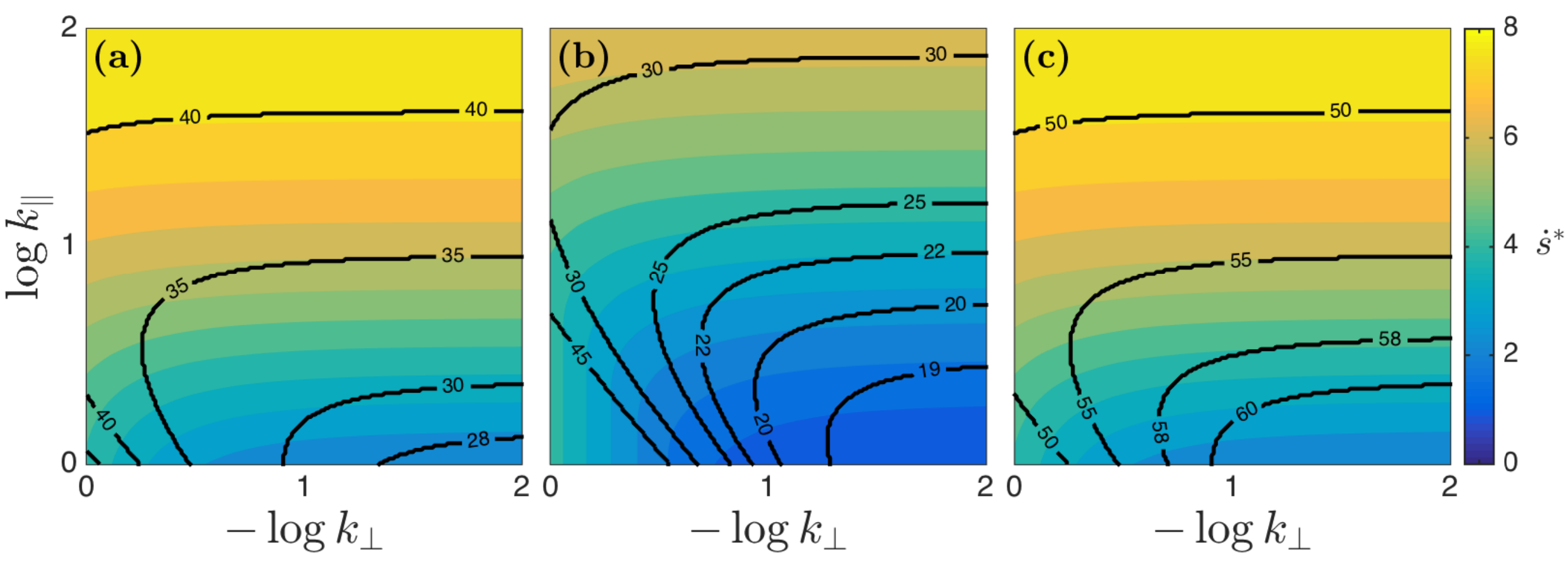}
  \caption{Contours of $\thetaMax$ (degrees) superimposed on a
    color-scale plot of $\sdotMax$ for
    $\kappa=1$. \textbf{(a)}~$\Theta=20^{\circ}$.
    \textbf{(b)}~$\Theta=45^{\circ}$.
    \textbf{(c)}~$\Theta=70^{\circ}$. }
  \label{fig:thetMax_ab}
\end{figure}

To search for anisotropy conditions conducive to formation of
low-angle bands, we map out the effect of anisotropy on $\thetaMax$.
Fig.~\ref{fig:thetMax_ab} displays contours of $\thetaMax$ and is
coloured according to the corresponding (maximum) growth rate
$\sdotMax$, for fixed $\kappa=1$. This is done in the space of
$\left(-\log\kperp,\log\kpara\right)$ because $\kperp$ ranges between
zero and one, while $\kpara$ is greater than or equal to one. For any
point in this space at which $\sdot$ has two maxima with equal rates,
we select the one at smaller $\theta$, corresponding to bands that are
more slowly rotated by the base-state flow. In these plots, the
lower-left corner corresponds to the isotropic case (where
$\thetaMax=45^{\circ}$) while the top-right corner corresponds to the
case with strong anisotropy.

Fig.~\ref{fig:thetMax_ab} demonstrates that $\thetaMax$ is most
effectively reduced by a sharp decrease of $\kperp$ with little or no
change in $\kpara$. Panel~(b) uses $\Theta=45^{\circ}$ and shows that
in this case, $\thetaMax$ can be lowered to below $20^{\circ}$ by
imposing $\kappa\sim1$, $\kpara\sim1$, and $\kperp\to0$. Under these
conditions the predicted angles are consistent with bands observed in
experiments, however $\sdot$ is low and so rotation due to the
background flow is likely to result in bands observed at higher angles
than the angle of maximal instantaneous growth rate. Panels~(a) and
(c), however, show that for $\Theta=20^{\circ}$ and
$\Theta=70^{\circ}$, $\thetaMax$ cannot be lowered to angles
consistent with the experimental band angles.

All three panels of Fig.~\ref{fig:thetMax_ab} show that away from
the origin, the maximal growth rate $\sdotMax$ increases with $\kpara$
and is nearly independent of $\kperp$. Close to the origin, where the
permeability is approximately isotropic, the growth rate does depend
on $\kperp$ (see especially panel~(b)). For bands to grow, the melt
segregation velocity must have a component normal to the perturbation
wavefronts. And as we have seen above, growth rate is sensitive to the
orientation of perturbations with respect to MPO and to the
perturbation wavelength relative to the compaction length. A simple
heuristic that combines these dependencies is the non-dimensional
compaction length in the direction normal to perturbation wavefronts
of the fastest-growing perturbation.  We define this as
\begin{displaymath}
  \dperp \equiv \sqrt{\left\langle\kap^*, \mathbf{A}\kap^*\right\rangle} 
  = \kappa\sqrt{\kpara - (\kpara-\kperp)\sin^2(\thetaMax+\Theta)}.
\end{displaymath}
This definition requires that when MPO is parallel to maximally
growing perturbations ($\thetaMax+\Theta=90^\circ$ or $270^\circ$),
the compaction length across bands is controlled by $\sqrt{\kperp}$,
whereas when MPO is normal to these perturbations, the compaction
length is controlled by $\sqrt{\kpara}$.  Returning to
Fig.~\ref{fig:thetMax_ab}, far from the origin we have
$\kpara\gg\kperp$ and $\sin^2(\thetaMax+\Theta)\neq1$, so $\dperp$
(and hence $\sdotMax$) is primarily dependent on $\kpara$. Near the
origin, $\kpara\approx\kperp\approx 1$ and so $\dperp$ depends on both
permeability parameters.  In panel~(b) near the origin, we have the
special case of $\thetaMax=\Theta=45^{\circ}$; hence
$\dperp = \sqrt{\kperp}$ so $\sdotMax$ depends only on $\kperp$ here.

\section{Numerical simulations}
\label{sec:numerical}

The linearised stability analysis presented above is valid at
infinitessimal strains and only for the plane-wave perturbations
considered. Numerical solutions discussed in this section allow us to
verify and extend the linearised analysis.  We solve the
system~\eqref{eq:nondimensional_conservation} with permeability
tensor~\eqref{eq:permeability1} and shear and compaction viscosity as
specified above in~\S{\ref{sec:governing}}.  The domain is rectangular
with height $H$ and boundary conditions of tangengial velocity
$\vvec = \pm \tfrac{1}{2}\dot{\gamma}H \hat{\xvec}$ (dimensional;
positive velocity on the top boundary) and impermeability
$\hat{\yvec}\cdot\permtensor=\boldsymbol{0}$ imposed on the top and
bottom boundaries (the domain is periodic in the
$x$~direction). Spatial derivatives are discretised on a Cartesian,
fully staggered grid with square cells. A semi-implicit,
Crank-Nicolson scheme is used to discretise time, and the hyperbolic
equation for porosity evolution is solved separately from the elliptic
system in a Picard loop with two iterations at each time-step. The
solutions are obtained in the context of the Portable, Extensible
Toolkit for Scientific Computation \citep[PETSc,][]{petsc-homepage,
  petsc-manual, katz07}.  Full details and references are provided by
\cite{katz13}.

\begin{figure}[ht]
  \centering
  \includegraphics[width=16cm]{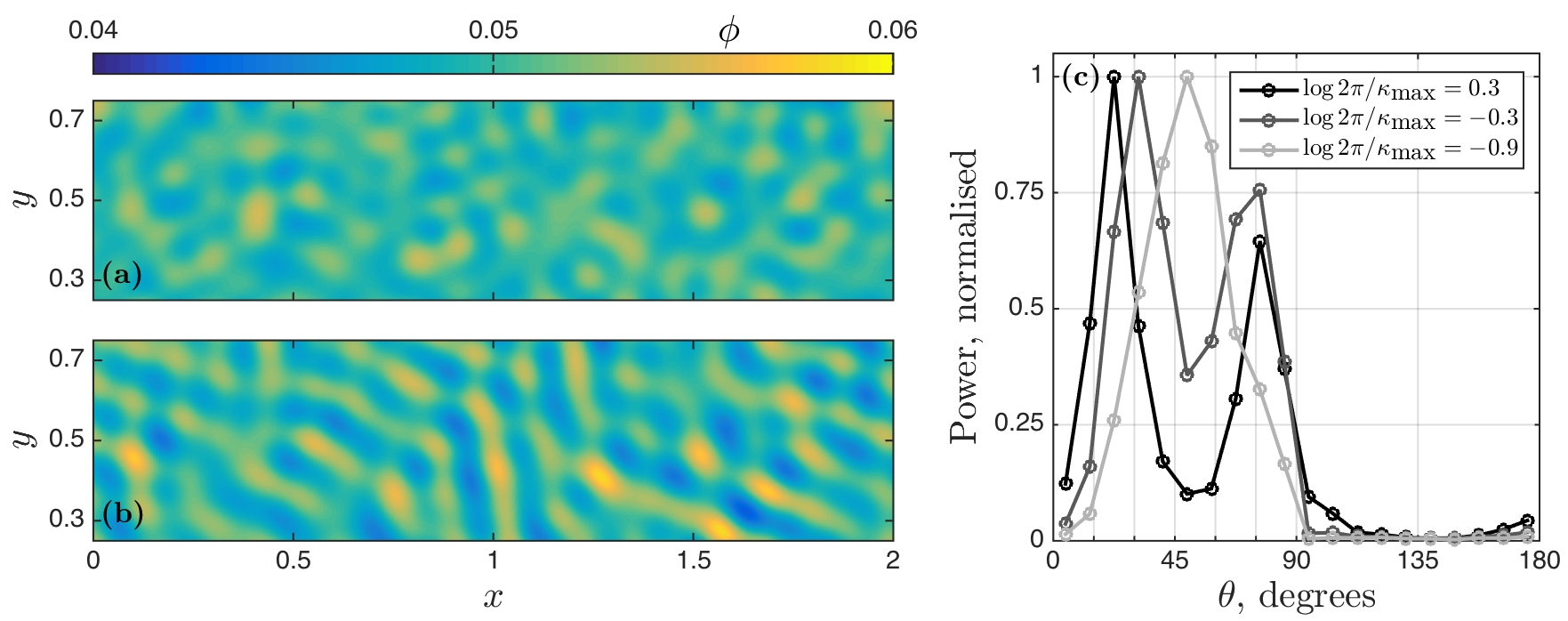}
  \caption{Results from numerical solutions
    of~\eqref{eq:nondimensional_conservation} with permeability
    tensor~\eqref{eq:permeability1} on a 1200$\times$600
    grid. Parameter values are $\kperp=0.005$, $\kpara=2$,
    $\Theta=45^\circ$. The coordinate axes $(x,y)$ have been rescaled
    with $\delta/H$, where $\delta$ is the compaction length and $H$
    is the domain height. \textbf{(a)} The initial porosity field with
    mean $\phi_0=0.05$ and random perturbation of amplitude
    $\smallpar=\tfrac{1}{10}\phi_0$.  The perturbation is white noise,
    filtered to remove wavelengths smaller than $1/10^\text{th}$ the
    domain height. This field is used to initialise all
    simulations. \textbf{(b)} The porosity field at strain
    $\gamma=0.4$ from a simulation with the compaction length chosen
    as $\delta = 0.05 H$ such that
    $\log 2\pi/\kappa_\text{max} = 0.3$.  \textbf{(c)} Spectral power
    binned by perturbation wavefront-angle $\theta$ to the shear plane
    \citep[after][]{katz06} for the porosity field at $\gamma=0.4$
    from simulations with values of $\delta$. Curves are normalised
    to have unit maximum.}
  \label{fig:numerical}
\end{figure}

Numerical simulations of band formation with isotropic permeability
are documented in the literature \citep[e.g.][]{katz06, katz13} and
not repeated here.  Instead, we consider simulations in which the
anisotropy is fixed at $\Theta=45^\circ$, $\kperp=0.005$, and
$\kpara=2$; we vary the ratio of the smallest perturbation wavelength
to the compaction length. All simulations are initialised with the
identical porosity field, shown in Fig.~\ref{fig:numerical}(a). This
field is constructed as
$\phi(\xvec,t=0) = \phi_0 + \smallpar\phi_1(\xvec)$, where
$\phi_0=0.05$ and $\epsilon=0.005$. $\phi_1(\xvec)$ is a smooth,
random field with unit amplitude, generated by filtering grid-scale
white noise to remove variation at wavelengths below $H/10$.  Hence
the smallest perturbation wavelength is
$2\pi/\kappa_\text{max} = H/(10\delta)$, in non-dimensional terms that
correspond to the abscissa in Fig.~\ref{fig:wavelength_effect}. A
suite of simulations at different values of compaction length $\delta$
was run to compare the distribution of angles for a broadband
perturbation.

If the prediction from the linearised stability analysis holds, we
expect simulations to produce bands at low ($\sim$$20^\circ$)
and high ($\sim$$70^\circ$)
angles when the minimum perturbation wavelength is greater than the
compaction length, $H/(10\delta)\gtrsim 1$.  In contrast, when the
minimum perturbation wavelength is smaller than the compaction length,
we expect a transition to $\theta \gtrsim 45^\circ$.  These
systematics are discernable in Fig.~\ref{fig:wavelength_effect}.

Fig.~\ref{fig:numerical}(b) shows the porosity field from a simulation
with $\delta=0.05H$ (and thus $\log 2\pi/\kappa_\text{max} = 0.3$)
after a strain of $\gamma=0.4$.  The porosity field, which was
initially as shown in panel~(a) of the same figure, has evolved to
form low- and high-angle porosity bands.  The angles of these features
are characterised in the power spectrum shown in panel~(c). Here,
spectral power from a two-dimensional Fourier transform is binned
according to the angle that wavefronts make with the shear plane.  The
black curve represents the case of perturbations with a minimum
wavelength that is larger than the compaction length. It is peaked at
$\sim$$22^\circ$
to the shear plane, with a secondary peak at $\sim$$75^\circ$.
This asymmetry is in contrast with the equal heights of growth-rate
peaks in Fig.~\ref{fig:VariousTheta} for $\Theta=45^\circ$.  However,
the relatively smaller height of the high-angle peak is expected
because under simple-shear deformation, high-angle features are
rotated more rapidly than low-angle features; this means that they
spend less time in optimal orientation for growth \citep{katz06}.

Fig.~\ref{fig:numerical}(c) compares the angular spectra of
simulations with three different values of
$\log 2\pi/\kappa_\text{max}$.  As the minimum wavelength of
perturbation decreases from longer than to shorter than the compaction
length, the dominant band orientation rotates to higher angle. For
$\log 2\pi/\kappa_\text{max} \sim -1$, the low- and high-angle peaks
have merged into a single peak at about $45^\circ$ (slightly higher
due to rotation).  This confirms the prediction of linearised analysis
over finite strains.

\section{Discussion and conclusion}
\label{sec:discussion}

The foregoing theory and results extend the equations of two-phase,
magma/mantle dynamics to include anisotropic permeability.
Anisotropic permeability is likely to arise as a consequence of
melt-preferred orientation, the coherent alignment of melt pockets
between solid grains. This alignment may be a consequence of finite
strain or of deviatoric stress, or some combination of the two.
Although we don't model details of the two-phase microstructure
\citep{hersum2008}, we assume that MPO is a consequence of deviatoric
stress, develops instantaneously, and is thus active at $t=0$
\citep{takei09a}. We postulate a simple anisotropy tensor with two
eigenvalues representing the MPO-parallel and MPO-perpendicular
permeabilities.  An angle between the shear plane and the
perpendicular to MPO then becomes the third parameter in the
anisotropy model.

To understand the behaviour of the system of governing equations and
constitutive laws, we perform linearised stability analysis and obtain
numerical solutions.  We explore the $(\kpara,\kperp,\Theta)$
parameter space of the anisotropy model with a particular focus on
$\Theta=45^\circ$, representing MPO aligned perpendicular to the
maximum tensile direction of the deviatoric stress tensor.

For perturbations with wavelengths approximately equal to the
compaction length, anisotropic permeability modifies the angular
dependence of the perturbation growth rate. It suppresses the growth
of high-porosity bands when those bands are in or near alignment with
the MPO (high-permeability) direction. This is because melt that
segregates into MPO-parallel bands must overcome the reduced
permeability perpendicular to MPO.  For $\Theta=45^\circ$, MPO is
aligned normal to the direction of maximum tensile stress, and hence
the reduced permeability resists band growth at precisely the angle
that would be optimal in the absence of anisotropy.  In this case, the
symmetric peaks in growth rate arise at angles where the product of
band-perpendicular tension and band-perpendicular permeability is
maximised. This angle selection, however, depends on the wavelength of
perturbation. Wavelengths much longer than the compaction length do
not generate sufficient pressure gradients to cause melt segregation
and band growth; wavelengths much shorter than the compaction length
are unhindered by the extremely short pathways of melt segregation
(over which they experience the anisotropy), and hence grow optimally
at $45^\circ$.

Of particular interest in this exploration is the question of whether
anisotropy of permeability can give rise to bands of high porosity
that are oriented at low angle ($\lesssim 20^\circ$) to the shear
plane, as observed in laboratory experiments. Our analysis predicts
that it is possible, under certain conditions
(i.e.~$\Theta\sim 45^\circ,\,2\pi/\kappa\sim 1,\,\kpara/\kperp\gg1$),
to obtain bands oriented at the low angles that are consistent with
experiments. These are fairly restrictive conditions.  The requirement
that initial porosity perturbations be of a wavelength that is greater
than about one compaction length seems especially significant. Also,
the angle of MPO that is typically measured under relevant
experimental conditions is closer to $\Theta=30^\circ$, as opposed to
$45^\circ$, though there is no accepted explanation for this.

On the other hand, there are arguments for the conditions above to be
met, in which case anisotropic permeability might play a role in
shaping and orienting the high porosity bands. Firstly, bands observed
in experiments tend not to have wavelengths of a much smaller scale
than the compaction length \citep{holtzman07}. This suggest
that either the initial porosity variations are indeed concentrated at
a relatively large scale or, as briefly considered above, the
non-linear system is regularised by a process that occurs at a scale
close to the compaction length. This could occur by surface tension
acting over a diffuse interface \citep{bercovici15} or by diffusion of
chemical components \citep{takei09d}.  Secondly, the argument given by
\cite{butler10} for the growth of band wavelengths suggests that the
porosity variations would evolve kinematically to a length scale
comparable to the compaction length and hence the system would become
susceptible to the effects of anisotropic permeability.

Although anisotropic permeability in the form discussed here may not
independently explain the low angle of porosity bands in laboratory
experiments, it may still be relevant in shaping the dynamics of those
experiments, or of partially molten rocks more generally.  Anisotropy
of permeability may, for example, affect the trajectories of rising
magma beneath mid-ocean ridges \citep{phippsmorgan87} or melt
segregation in magma chambers \citep{bergantz95}; in industrial
processes, it has implications for macrosegregation during
solidification \citep[e.g.][]{yoo92}.  More work is therefore needed
to understand the consequences of anisotropy as well as the causes.
In this latter category: How can melt-preferred orientation be
quantitatively related to permeability?  And, even more fundamentally,
what is the energetic or mechanical cause for melt-preferred
orientation?

\paragraph{Acknowledgements} The authors thank S.~Butler and an
anonymous reviewer for insightful comments. The research leading to
these results has received funding from the European Research Council
(ERC) under the European Union’s Seventh Framework Programme
(FP7/2007–2013)/ERC grant agreement 279925. JT-W was supported for
summer 2014 by as Research Experience Placement grant from the Natural
Environment Research Council of the Research Councils UK. RFK is
grateful for the support of the Leverhulme Trust. Numerical solutions
were computed at the Advanced Research Computing facility of the
University of Oxford.

\appendixpage
\appendix
\numberwithin{equation}{section}
\section{Diagonality of permeability tensor}
\label{sec:diagonal}
\begin{figure}[h]
  \centering
  \includegraphics[width=.5\linewidth]{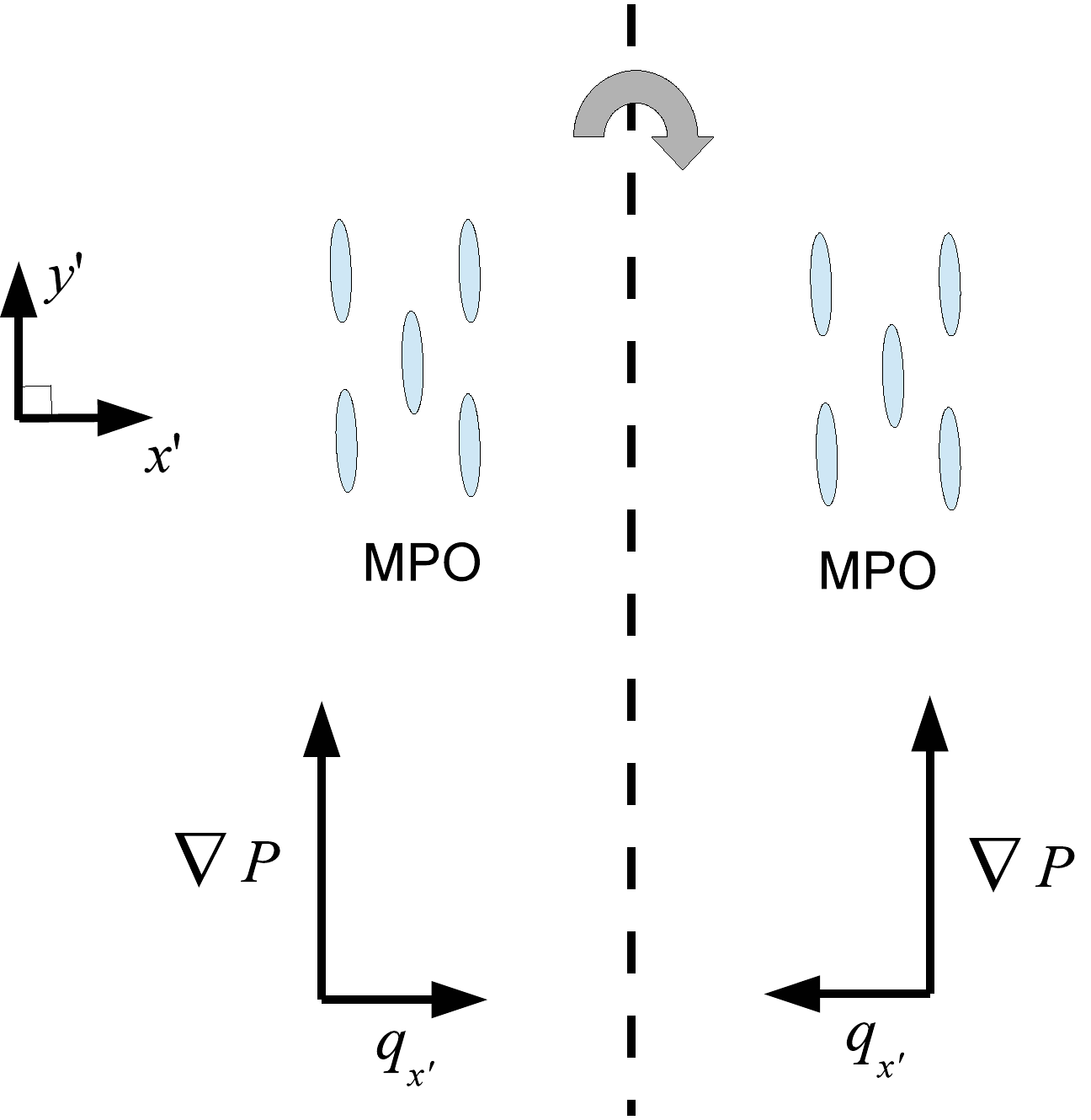}
  \caption{Reflection of MPO in $y'$--axis}
  \label{fig:Reflection}
\end{figure}

To understand why the permeability tensor $\mathbf{A}$ must be
diagonal when expressed in a coordinate system aligned with MPO,
consider equation \eqref{Darcy} with uniform porosity $\phi=\phi_0$
and permeability given by equation~\eqref{eq:perm},
\begin{equation}
  \label{eq:Darcy_again}
  \phi_0\left(\vs-\vl\right)=\frac{K_0}{\mu}\mathbf{A}\Grad p^L.
\end{equation}
We now adopt the primed coordinate system of Figs.~\ref{fig:Shear} and
\ref{fig:Reflection} that is aligned with the melt-preferred
orientation. Making no assumptions about the form of the permeability
tensor, equation~\eqref{eq:Darcy_again} becomes
\begin{align}
  \label{segeq}
  \boldsymbol{q}=\frac{K_0}{\mu}\left(
  \begin{array}{cc} \kperp & k_{\times_1}  \\
    k_{\times_2} & \kpara   \end{array} \right)\Grad p^L,
\end{align}
where $\boldsymbol{q}=\phi_0(\vs-\vl)$ is the segregation flux. For a
unit pressure gradient, $\Grad p^L=(0, 1)$, applied in the
$y'$-direction we have
\begin{equation}
  \label{eq:seg}
  \boldsymbol{q}=\frac{K_0}{\mu}(k_{\times_1}\hat{\xvec}' 
  + \kpara\hat{\yvec}')
\end{equation}
with an $x'$-component $q_{x'}$ proportional to $k_{\times_1}$. 

If we reflect in an axis parallel to the $y'$-direction, as shown in
Fig.~\ref{fig:Reflection}, the pressure gradient is unchanged but
the cross-MPO segregation $q_{x'}$ becomes proportional to
$-k_{\times_1}$. However, the melt-pocket structure in
Fig.~\ref{fig:Shear} is unchanged and so the permeability tensor
should not change. Hence the segregation obtained from \eqref{segeq}
is still given by equation~\eqref{eq:seg} and we have
$k_{\times_1}=-k_{\times_1}=0$, and by a similar argument
$k_{\times_2}=0$.

\section{Solving the linearised equations for $\sdot$}
\label{sec:solve}
Because the system of equations that governs the perturbation
quantities is linear, we are able to relate the other variables to
$\phi_1$ via unknown coefficients,
\begin{equation}
  \vvec^{(1)}=\tilde{\vvec}\phi_1, \hspace{4mm} P_1=\tilde{P}\phi_1,
\end{equation}
The order-$\smallpar$ balance derived from equation
\eqref{nondimconseq1} then gives
\begin{equation}
  \label{eq:pdiffphi}
  \pdiff{\phi_1}{t} = \left(1-\phi_0\right)\tilde{\vvec}
  \cdot\Grad\phi_1-\vvec^{(0)}\cdot\Grad\phi_1.
\end{equation}
Expanding time and space derivatives using~\eqref{perturbation} and
rearranging, equation~\eqref{eq:pdiffphi} becomes an equation for the
growth rate
\begin{equation}
  \label{Growth rate}
  \sdot=\left(1-\phi_0\right)i\kap\cdot\tilde{\vvec} =
  \left(1-\phi_0\right)\cmp_1/\phi_1.
\end{equation}

Substituting $\vvec^{(0)}=(Y,0)$, $\vvec^{(1)}=\tilde{\vvec}\phi_1$ and
$P_1=\tilde{P}\phi_1$ into the $x$ and $y$ components of
equation~\eqref{gradP} gives
\begin{subequations}
  \begin{align}
    \tilde{P}\kappa_x &= i\left[\left(\rxi+4/3\right)\kappa_x^2\tilde{V_x}
                        +\left(\rxi+1/3\right)\kappa_x\kappa_y\tilde{V_y} + 
                        \kappa_y^2\tilde{V_x}\right]-\lambda\kappa_y, \\
    \tilde{P}\kappa_y &= i\left[\left(\rxi+4/3\right)\kappa_y^2\tilde{V_y} 
                        +\left(\rxi+1/3\right)\kappa_x\kappa_y\tilde{V_x} + 
                        \kappa_x^2\tilde{V_y}\right]-\lambda\kappa_x,
  \end{align}
\end{subequations}
which can be combined using $\kappa_x^2+\kappa_y^2=\kappa^2$
to obtain
\begin{align}
  \tilde{P}&=i\left(\rxi+4/3\right)\left(\kappa_x\tilde{V_x}
             +\kappa_y\tilde{V_y}\right)-2\frac{\kappa_x\kappa_y}{\kappa^2}\lambda \nonumber\\ 
           &=\left(\rxi+4/3\right)\cmp_1/\phi_1-\lambda\sin{2\theta}.
\end{align}
By substituting this into equation \eqref{compaction1} we obtain
\begin{align}
  \cmp_1&=\frac{1}{\rxi+4/3}\Div\left(i\tilde{P}\mathbf{A}\kap\phi_1\right) \nonumber\\
        &=-\frac{\tilde{P}}{\rxi+4/3}\left\langle\kap,\mathbf{A}
          \kap\right\rangle\phi_1 \nonumber\\
        &=-\left\langle\kap,\mathbf{A}\kap\right\rangle\cmp_1
          +\frac{\lambda\sin{2\theta}}{\rxi+4/3}\left\langle\kap, 
          \mathbf{A}\kap\right\rangle\phi_1,
\end{align}
where the inner product notation means
$\left\langle\kap, \mathbf{A}\kap\right\rangle =\sum_{i}\limits
\sum_{j}\limits\kappa_i A_{ij}\kappa_j$.
This provides an expression for $\cmp_1/\phi_1$ that, when substituted
into \eqref{Growth rate}, gives the growth rate
\begin{align} 
  \sdot = \lambda(\rxi+4/3)^{-1}\left(1-\phi_0\right)
  \sin{2\theta}\frac{\left\langle\kap,\mathbf{A}
  \kap\right\rangle}{1+\left\langle\kap,\mathbf{A}\kap\right\rangle}.
\end{align}

To understand the role of perturbation wavelength in controlling the
growth rate we consider the anisotropy factor
\begin{equation}
  F=\frac{\left\langle\kap,\mathbf{A}
      \kap\right\rangle}{1+\left\langle\kap,\mathbf{A}\kap\right\rangle}=
  \frac{\kappa^2\left[\kpara - (\kpara-\kperp)\sin^2(\theta+\Theta)\right]}
  {1+\kappa^2\left[\kpara - (\kpara-\kperp)\sin^2(\theta+\Theta)\right]}.
\end{equation}
A rough measure of the effectiveness of anisotropy of permeability in
controlling the angular dependence of the growth rate is the range of
variation of $F$.  The maximum and minimum values of $F$ over all
$(\theta+\Theta)$ are
\begin{equation}
  F_{\textrm{max}}=\frac{\kappa^2\kpara}{1+\kappa^2\kpara }, \hspace{0.5in}
  F_{\textrm{min}}=\frac{\kappa^2\kperp}{1+\kappa^2\kperp }.
\end{equation}
We seek the stationary point of $F_{\textrm{max}}-F_{\textrm{min}}$
with respect to wavenumber $\kappa$,
\begin{align}
  0 = \frac{\mathrm{d}}{\mathrm{d}\kappa^2}\left(F_{\textrm{max}}-F_{\textrm{min}}\right)
  &=\frac{\kpara}{\left(1+\kpara\kappa^2\right)^2}-\frac{\kperp}{\left(1+\kperp\kappa^2\right)^2}
  \\ \implies \kappa^2 &= \frac{1}{\sqrt{\kpara\kperp}}.
\end{align}
Taking a second derivative gives
\begin{align}
  \frac{\mathrm{d}^2}{\mathrm{d}(\kappa^2)^2}\left(F_{\textrm{max}}-F_{\textrm{min}}\right)
  &= \frac{2\kperp^2}{\left(1+\kperp\kappa^2\right)^3}-\frac{2\kpara^2}{\left(1+\kpara\kappa^2\right)^3},
\end{align}
which, evaluated at $\kappa^2=1/\sqrt{\kpara\kperp}$, gives
\begin{align}
  \frac{2\left(\sqrt{\kperp\kpara}\right)^3}{\left(\sqrt{\kperp}+\sqrt{\kpara}\right)^3}
  \left(\sqrt{\kperp}-\sqrt{\kpara}\right)<0.
\end{align}
This result shows that this stationary point is a maximum in the total
variation of $F$. Using the example of $\kpara=2$, $\kperp=0.005$ as
in the numerical simulations, we find that the perturbation most
affected by anisotropy has non-dimensional wavelength
$2\pi/\kappa=2\pi/\sqrt{10}=1.987...\sim1$, as expected.

\bibliographystyle{abbrvnat} 
\bibliography{manuscript.bib}

\end{document}